# Segmentation of skin lesions and their attributes using Generative Adversarial Networks

Cristian L.
Mechatronics Engineering Faculty
Universidad Nacional de Ingeniería
Lima, Peru 15333
`clazoq@uni.pe`

January 30, 2021

## Abstract

This work is about the semantic segmentation of skin lesion boundary and their attributes using Image-to-Image Translation with Conditional Adversarial Nets. Melanoma is a type of skin cancer that can be cured if detected in time. Segmentation into dermoscopic images is an essential procedure for computer-assisted diagnosis due to its existing artifacts typical of skin images. To alleviate the image annotation process, we propose to use a modified Pix2Pix network. The discriminator network learns the mapping from a dermal image as an input and a mask image of six channels as an output. Likewise, the discriminative network output called PatchGAN is varied for one channel and six output channels. The photos used come from the 2018 ISIC Challenge, where 500 photographs are used with their respective semantic map, divided int 75% for training and 35% for testing. Obtaining for 100 training epochs high Jaccard indices for all attributes of the segmentation map.

## 1 Introduction

According to [1], ], early detection of cancer is essential because skin cancer is one of the most common and fast-growing cancers globally. Early treatment could save many lives; for that reason, the automatization of segmentation of skin melanomas is a fundamental basis in skin cancer detection. The photographs have artifacts that decrease the cancer detection model's performance, as indicated in [2]. These artifacts are typical of the data acquisition process and the dermatoscopy used, such as hair, bubble, and ruler.

This process of segmentation of skin lesions would allow better performance of classification models, and people could access the treatment in due time. The work consists of three main parts: first, preparing the data to properly use the convolutional neural network. The boundary lesion masks and the attributes form a matrix of sin channels; this matrix is stored in two images of three channels each to be used later. The second part is the implementation of the proposed architecture. The Pix2Pixarchitecture is used to modify the discriminator and the generator for semantic segmentation tasks. All algorithms have been implemented in python using TensorFlow. Finally, predictions for test data are evaluated using the Jaccard index as a metric.Code is available at `https://github.com/CristianLazoQuispe/skin-lesion-segmentation-using-pix2pix.git`

## 2  Related work

This section summarizes some of the most relevant articles taken as reference for the development of this work.

The Pix2Pix algorithm [3], is one of the first successful general images to image translation algorithms. Mostly used to generate realistic images (or any domain), and for coloring and super-resolution tasks. The architecture is composed mainly by two networks: discriminator and one generated. The semantic segmentation task would usually pass through a parametric function, and the difference in the resulting image and the ground truth output would be used to update the network weights. However, this loss function's design with standard distance measurements such as L1 and L2 will not capture many of the important distinguishing features between these images. For this reason, it is proposed to use opposing generative networks to take advantage of the discriminator of the pix2pix architecture. This work combines the Conditional-Adversarial Loss (Generator versus Discriminator) and the L1 loss function to obtain a better cost function that can better abstract if the target image resembles the generated image. The Conditional-Adversarial Loss(Generator versus Discriminator)format is as showed in equation 1. In addition, the L1 loss function previously mentioned is shown in equation 2. Combining these functions results in one loss function. Since the game-theoretic approach is taken, the target function is represented as a minmax function. The discriminator tries to maximize the target function so that we can make gradient ascents over the target function. The generator tries to minimize the objective function; therefore, we can perform gradient descent on the objective function. By alternating between gradient ascent and descent, the network can be trained. The format of the final cost function is shown in equation 3.

$$\mathcal{L}_{cGAN}(G, D) = \mathbb{E}_{x,y}[log D(x, y)] + \mathbb{E}_{x,z}[log(1 - D(x, G(x, z)))] \quad (1)$$

$$\mathcal{L}_{L1}(G) = \mathbb{E}_{x,y,z}[||y - G(x, z)||_1] \quad (2)$$

$$G^* = arg \min_{G} \max_{D} \mathcal{L}_{cGAN}(G, D) + \lambda \mathcal{L}_{L1}(G) \quad (3)$$

In [4], they use the 2018 ISIC Challenge dataset as input a semantic mask and output an image of a cutaneous lesion. As a data augmentation method using the Pix2pixHD algorithm, whose input is the segmented image with its attributes to generate a better synthetic image.

In [5], they propose a Generative Adversarial Network to perform segmentation using the data set ISBI 2017 and the dataset ISIC 2018. The validation and testing sets contain 100 and 1,000images for the data ISIC 2018, where they obtained a 0.784 jaccard index for the test data.

## 3  Problematics

In 2017 [6], 10650 cases of skin cancer were registered in Peru, and 63.8% of cancers were detected when the patients already presented the symptoms caused by cancer. These figures are alarming for Peru because a late diagnosis of cancer in the worst case can lead to death.

The original data set of the ISIC 2018 [7] competition consists of 2594 dermoscopic images available to the public to analyze skin lesions. The input data are dermoscopic lesion images in jpeg format. All lesion images are named using the scheme ISIC_$image\_id$.jpg, where $image\_id$ a7-digit unique identifier. The dermatoscopic lesion images have their respective lesion boundary segmentation and their attributes: pigment network, negative network, streaks, milia-like cysts, and globules, as shown in Fig.1. However, certain melanoma images do not present all the attributes, as shown in Fig.1a and Fig.1b. The model needs to learn to recognize when these attributes are present; thus, the output is generated with six matrices that represent each mask. These masks will be the output of our Generative Adversarial Network.

As mentioned in [8], images of lesions were obtained with a variety of dermatoscopic types, from all anatomical sites (excluding mucosa and nails), from a historical sample of patients presented for skin cancer detection, from several different institutions. Every lesion image contains exactly one primary lesion; other fiducial markers, smaller secondary lesions, or other pigmented regions may be neglected.

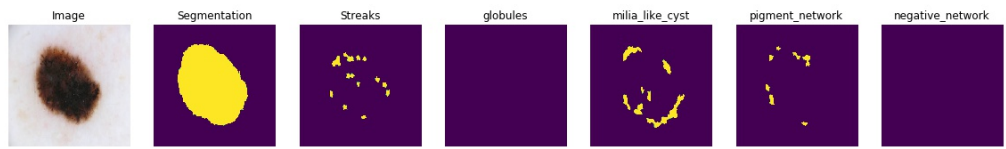

(a) Example of lesion image called ISIC_0000013

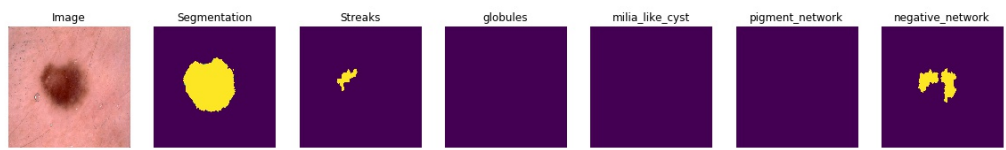

(b) Example of lesion image called ISIC_0014867

Figure 1: Example of two images of melanoma with different attributes. The image (a) only has the negative_network attribute and The image (b) only has the milialike cysts attribute.

## 4 Proposed Method

### 4.1 Data preparation

First, the appropriate data set is prepared for later use of the modified model Pix2pix. The output of our network will be a 6-channel image, each channel representing an attribute in a binary mask, as shown in figure 2. First, we divide the images of 6 channels into 2 pictures of 3 channels to save them as normal images in .PNG format. The concatenated image of these two arrays is used as input to the neural network. This technique is used to take advantage of the efficient data reading functions with TensorFlow to form a dataset. In Fig.2a and 2c, lesion boundary segmentation, pigment network, and negative network and images are concatenated into a single image and stored in a folder. As shown in Figure 2b and 2d, streaks, milia-like cysts, and globules images are concatenated into a single image and stored in a folder.

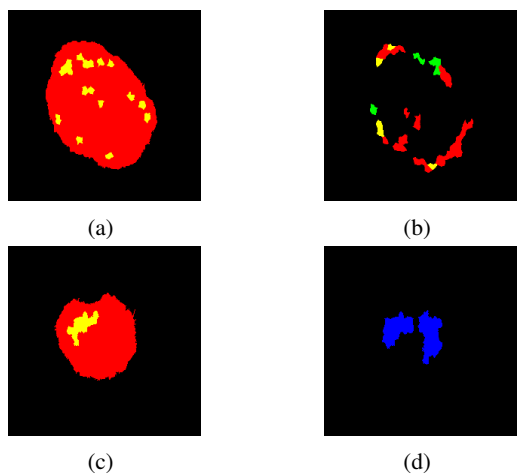

Figure 2: In (a) and (b) are shown the attributes of the lesion ISIC_0000013, in (c) and (d) of the lesion ISIC_0014867.The colors shown are only a proper representation of a 3-channel RGB image that depends on the combination of these.

## 4.2 Implementation algorithm

The Work [3] is taken as a reference. Two main changes are made. First, the generative network's output is modified to obtain an image of 6 channels; each channel represents a mask of the lesion. Figure 3 shows an example of the generative network output, being this an image of 256x256x6. In this way, the generative adversarial network is used to learn to map from a skin lesion image as input to a set of masks as outputs. Therefore, you have the advantage of generating all 6 segmentations simultaneously using only one neural network. Finally, as mentioned above, the Pix2pix network uses the neural network called PatchGan taken as reference from [9]. This architecture is used as a discriminator to improve our cost function and to be able to use an intelligent discriminator that learns to differentiate if the masks generated are similar to the original masks. Two tests are performed, varying the number of output channels of the PatchGannetwork. In [2], the authors mention that the output is a 30x30 image where each pixel value (0 to 1)represents how credible the corresponding section of the unknown image is. In the implemented network the tests are used for an output of 30x30x1 and 30x30x6. In the first test, each pixel of this30x30 image corresponds to a 256x256x6 generated image's credibility, making a general analysis of the grouped regions. As we have six output channels in the second test, each 30x30 matrix represents a 256x256 mask's credibility. We wanted to test if doing this variation obtains better results than using only one. Figure 3 shows an example of the output of the discriminating network.

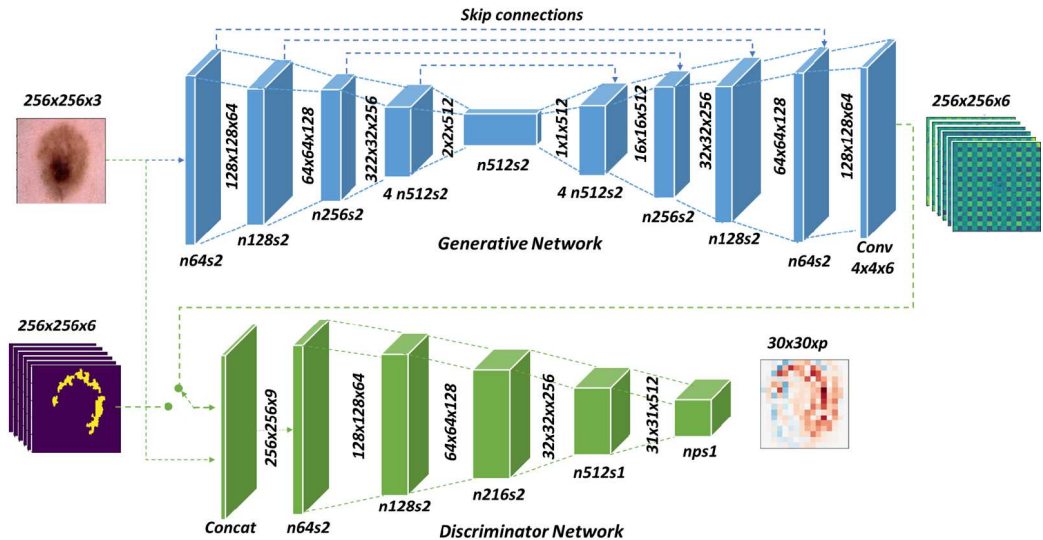

Figure 3: The architecture proposed.

The training method should take into consideration; two steps must be taken: train the discriminator and train the generator. When calculating the gradients, the discriminator is used twice, as shown in figure 3 at the discriminating network entrance.

First, the generator generates an output image. The discriminator looks at the input/destination pair and the input/output pair and generates his guess about how realistic they look. The discriminator weight vector is then adjusted based on the input/output torque classification error and the input/target torque. The generator weights are adjusted according to the discriminator output and the difference between the output and the target image.

It is essential to identify that when the generator is trained at the discriminator output, the gradient is actually calculated through the discriminator, which means that while the discriminator is improving, the generator is being trained to beat the discriminator. Suppose the discriminator is good at his job and the generator is able to learn the correct mapping function through gradient descent. In that case, you should get generated results that could be as real as the original masks.

Two types of tests are performed. First, only the skin lesion is segmented. After this, the segmentation is done, including the attributes to compare the proposed method's efficiency. Only 500 images are used, divided into 75% for training and 25% for testing.

## 5 Results

Two training tests are performed by changing the number of channels of the output of the discriminative network.

For the first segmentation task it was tested having only one output channel of the discriminator. Using a training input of 100 times the following jaccard indices were obtained as shown in table 1.

Table 1: Results for the first training

| Atributte | Jaccard index |
|---|---|
| Lesion boundary | 0.85 |
| pigment network | 0.82 |
| negative network | 0.81 |
| streaks | 0.75 |
| milia-like cysts | 0.76 |
| globules | 0.83 |

For the second segmentation task it was tested having six output channel of the discriminator. Using a training input of 100 times the following jaccard indices were obtained as shown in table 2.

Table 2: Results for the second training

| Atributte | Jaccard index |
|---|---|
| Lesion boundary | 0.81 |
| pigment network | 0.79 |
| negative network | 0.80 |
| streaks | 0.78 |
| milia-like cysts | 0.77 |
| globules | 0.81 |

## 6 Future work

The algorithm works appropriately, but the main objective is to improve the efficiency of melanoma detection algorithms. For this reason, it is proposed to continue working implementing the classification part. The Generative Adversarial Networks will continue to be used, focusing the main task not on the generator but instead on the discriminator to classify melanoma or non-melanoma.

## 7 Conclusions

This article shows an efficient system for the task of segmentation, using few input images. Using only one output channel for the discriminator is better than using six, agreeing as mentioned in the paper of Pix2pix, putting more channels on the output imposes more restrictions that encourage sharp high-frequency detail. Likewise, the generative network works well mapping the original lesion image to the segmented mask set. The masks tend to be more similar to the training images, competing for the generator network with the discriminator network. Finally, since the algorithm, developed uses free software, its use is accessible, and there would be no restriction on its use.